\documentclass{PoS}

\usepackage{axodraw}

\title{Two-loop electroweak Sudakov logarithms for massive fermion scattering}

\ShortTitle{Two-loop electroweak Sudakov logarithms for massive fermion scattering}

\author{Ansgar Denner, Bernd Jantzen\\
Paul Scherrer Institut, CH-5232 Villigen PSI, Switzerland\\
E-mail: \email{Ansgar.Denner@psi.ch}, \email{physics@bernd-jantzen.de}}

\author{\speaker{Stefano Pozzorini}\\
Max-Planck-Institut f\"ur Physik, F\"ohringer Ring 6, D-80805 M\"unchen, Germany\\
E-mail: \email{pozzorin@mppmu.mpg.de}}

\abstract{ We study the asymptotic behaviour of two-loop electroweak
  corrections at energies $Q^2\gg \MW^2$, where logarithms of the type
  $\ln(Q^2/\MW^2)$ become dominant.
  The calculation of the leading and next-to-leading logarithmic terms for
  massless and massive fermion-scattering processes is summarized.
  The derivations are performed diagrammatically within the spontaneously
  broken electroweak theory.
  We find that the soft and collinear singularities resulting from photons can
  be factorized into a QED-like term and that, up to logarithms of the Z--W
  mass ratio, the effects of symmetry breaking cancel.
  This result supports resummation prescriptions that are based on a symmetric
  $\SU(2)\times\U(1)$ theory matched with QED at the electroweak scale.}

\FullConference{8th International Symposium on Radiative Corrections \\
		 October 1-5, 2007\\
		 Florence, Italy}

\def\beq{\begin{equation}}
\def\eeq{\end{equation}}
\def\beqar{\begin{eqnarray}}
\def\eeqar{\end{eqnarray}}
\def\barr#1{\begin{array}{#1}}
\def\earr{\end{array}}
\def\bfi{\begin{figure}}
\def\efi{\end{figure}}
\def\btab{\begin{table}}
\def\etab{\end{table}}
\def\bce{\begin{center}}
\def\ece{\end{center}}
\def\nn{\nonumber}

\def\text{\textstyle}

\def\de{\delta}
\def\teps{\varepsilon}
\def\veps{\epsilon}

\def\refeq#1{\mbox{(\ref{#1})}}

\def\citere#1{\mbox{Ref.~\cite{#1}}}
\def\citeres#1{\mbox{Refs.~\cite{#1}}}

\def\solid{\raise.9mm\hbox{\protect\rule{1.1cm}{.2mm}}}
\def\dash{\raise.9mm\hbox{\protect\rule{2mm}{.2mm}}\hspace*{1mm}}

\newcommand{\order}{\mathcal{O}}

\newcommand{\lsim}
{\mathrel{\raisebox{-.3em}{$\stackrel{\displaystyle <}{\sim}$}}}

\def\asymp#1%
{\mathrel{\raisebox{-.4em}{$\widetilde{\scriptstyle #1}$}}}

\def\Nequal#1%
{\mathrel{\raisebox{-.5em}{$\stackrel{=}{\scriptstyle\rm#1}$}}}

\newcommand{\TeV}{\unskip\,\mathrm{TeV}}

\def\mathswitchr#1{\relax\ifmmode{\mathrm{#1}}\else$\mathrm{#1}$\fi}

\newcommand{\PW}{\mathswitchr W}
\newcommand{\PZ}{\mathswitchr Z}

\newcommand{\Pb}{\mathswitchr b}

\newcommand{\Pt}{\mathswitchr t}

\def\mathswitch#1{\relax\ifmmode#1\else$#1$\fi}

\newcommand{\MW}{\mathswitch {M_\PW}}

\newcommand{\MZ}{\mathswitch {M_\PZ}}

\newcommand{\Mt}{\mathswitch {m_\Pt}}

\newcommand{\muD}{\mu_{\mathrm{D}}}
\newcommand{\muR}{\mu_\mathrm{R}}

\newcommand{\ew}{\mathrm{ew}}

\def\ie{i.e.\ }

\newcommand{\etal}{{\it et al.}}

\newcommand{\rd}{\mathrm{d}}

\newcommand{\ri}{\mathrm{i}}

\newcommand{\rL}{\mathrm{L}}
\newcommand{\rR}{\mathrm{R}}

\newcommand{\SU}{\mathrm{SU}}
\newcommand{\U}{\mathrm{U}}

\newcommand{\elm}{\mathrm{em}}
\newcommand{\sew}{\mathrm{sew}}

\newcommand{\M}{{\cal {M}}}

\newcommand{\gw}{g_2}

\begin{document}

\newcommand{\leg}[1]{\scriptstyle{#1}}

%%%%%%%%%%%%%%%%%%%%%%  Feynman diagrams %%%%%%%%%%%%%%%%%%%%%%
\newcommand{\wblob}{
\Vertex(-15.9138,3.75675){0.8}
\Vertex(-16.3512,-0.00758122){0.8}
\Vertex(-15.9138,-3.75675){0.8}
\Line(0.,0.)(-19.5,-9.75)
\Line(0.,0.)(-19.5,9.75)
\GCirc(0.,0.){10.9008}{1}
}
\newcommand{\blob}{
\Vertex(-15.9138,3.75675){0.8}
\Vertex(-16.3512,-0.00758122){0.8}
\Vertex(-15.9138,-3.75675){0.8}
\Line(0.,0.)(-19.5,-9.75)
\Line(0.,0.)(-19.5,9.75)
\GCirc(0.,0.){10.9008}{0.5}
}
\newcommand{\factblob}{\wblob \Text(0,0)[]{\scriptsize F}}
\newcommand{\nfactblob}{\wblob \Text(0,0)[]{\scriptsize N}}

\newcommand{\diaggeneric}{
\begin{picture}(120.,104.)(-28.,-52.)
\Gluon(-53.6656,0.)(0.,0.){3.2}{5}
\Line(0.,0.)(24.,12.)
\Line(24.,12.)(48.,24.)
\Line(24.,-12.)(0.,0.)
\Line(48.,-24.)(24.,-12.)
\GCirc(4.47214,0.){17}{0.8}
%external lines
\Text(84.,42.)[l]{$\leg{i}$}
\Text(84.,-42.)[l]{$\leg{j}$}
%gauge bosons
%inflowing lines
\Text(27.668,14.6453)[br]{}
\Text(28.,-14.)[tr]{}
\end{picture}
}

\newcommand{\diagone}[4]{
\begin{picture}(120.,104.)(-28.,-52.)
\Line(0.,0.)(48.,24.)
\Line(48.,24.)(80.,40.)
\Line(48.,-24.)(0.,0.)
\Line(80.,-40.)(48.,-24.)
\Photon(48.,24.)(48.,-24.){2.4}{3.5}
\Vertex(48.,24.){2}
\Vertex(48.,-24.){2}
%external lines
\Text(84.,42.)[l]{#1}
\Text(84.,-42.)[l]{#2}
%gauge bosons
\Text(53.6656,0.)[l]{#3}
%inflowing lines
\Text(27.668,14.6453)[br]{}
\Text(28.,-14.)[tr]{}
{#4}
\end{picture}
}

\newcommand{\diagnewb}[1]{
\begin{picture}(120.,104.)(-28.,-52.)
\Line(0.,0.)(30.,15.)
\Line(30.,-15.)(0.,0.)
{#1}
\end{picture}
}

\newcommand{\diagnewc}[1]{
\begin{picture}(120.,104.)(-28.,-52.)
\Line(0.,0.)(48.,24.)
\Line(48.,24.)(80.,40.)
\Line(48.,-24.)(0.,0.)
\Line(80.,-40.)(48.,-24.)
\Photon(48.,24.)(48.,-24.){2.4}{3.5}
\Vertex(48.,24.){2}
\Vertex(48.,-24.){2}
%external lines
\Text(84.,42.)[l]{$\leg{i}$}
\Text(84.,-42.)[l]{$\leg{j}$}
%gauge bosons
\Text(53.6656,0.)[l]{$\scriptscriptstyle{V_1}$}
%inflowing lines
\Text(27.668,14.6453)[br]{}
\Text(28.,-14.)[tr]{}
{#1}
\end{picture}
}
\newcommand{\diagnewd}[1]{
\begin{picture}(120.,104.)(-28.,-52.)
\Line(0.,0.)(80,0)
\PhotonArc(25.,-8.)(26.2488,17.7447,162.255){2}{5}
\Vertex(50.,0.){2}
\Text(84.,0)[l]{$\leg{i}$}
%gauge bosons
\Text(23,28.)[l]{{$\scriptstyle{V_\mu}$}}
%inflowing lines
\Text(27.668,14.6453)[br]{}
\Text(28.,-14.)[tr]{}
{#1}
\end{picture}
}

\newcommand{\diagnewe}[1]{
\begin{picture}(120.,104.)(-28.,-52.)
\Line(0.,0.)(80,0)
\Line(0.,0.)(50,50)
\PhotonArc(0,0)(48,0,45){2}{4}
\Vertex(48.,0.){2}
\Vertex(33.941,33.941){2}
\Text(84.,0)[l]{$\leg{i}$}
\Text(53.,53.)[l]{$\leg{j}$}
%gauge bosons
\Text(50,22.)[l]{{$\scriptstyle{V_\mu}$}}
%inflowing lines
\Text(27.668,14.6453)[br]{}
\Text(28.,-14.)[tr]{}
{#1}
\end{picture}
}

\newcommand{\diagoned}[1]{
\begin{picture}(120.,104.)(-28.,-52.)
\Line(0.,0.)(50,0)
\PhotonArc(50.,-50.)(70.71,90.,135.){2}{5}
\Text(54.,0)[l]{$\leg{i}$}
%gauge bosons
\Text(54,22.)[l]{{$\scriptstyle{V_\mu}$}}
%inflowing lines
\Text(27.668,14.6453)[br]{}
\Text(28.,-14.)[tr]{}
{#1}
\end{picture}
}

\newcommand{\diagonee}[1]{
\begin{picture}(120.,104.)(-28.,-52.)
\Line(0.,0.)(60,0)
\Line(0.,0.)(50,50)
\Vertex(25,25){2}
\Photon(25,25)(60,25){2}{3}
\Text(64.,0)[l]{$\leg{i}$}
\Text(53.,53.)[l]{$\leg{j}$}
%gauge bosons
\Text(64,22.)[l]{{$\scriptstyle{V_\mu}$}}
%inflowing lines
\Text(27.668,14.6453)[br]{}
\Text(28.,-14.)[tr]{}
{#1}
\end{picture}
}

\newcommand{\diagI}[5]{
\begin{picture}(120.,104.)(-28.,-52.)
\Line(0.,0.)(28.,14.)
\Line(28.,14.)(56.,28.)
\Line(56.,28.)(80.,40.)
\Line(28.,-14.)(0.,0.)
\Line(56.,-28.)(28.,-14.)
\Line(80.,-40.)(56.,-28.)
\Photon(56.,28.)(56.,-28.){2.4}{3.5}
\Photon(28.,14.)(28.,-14.){2.4}{2}
\Vertex(56.,28.){2}
\Vertex(56.,-28.){2}
\Vertex(28.,14.){2}
\Vertex(28.,-14.){2}
%external lines
\Text(84.,42.)[l]{#1}
\Text(84.,-42.)[l]{#2}
%gauge bosons
\Text(62.6099,0.)[l]{#3}
\Text(31.305,0.)[l]{#4}
%inflowing lines
\Text(41.5019,21.9679)[br]{}
\Text(42.,-21.)[tr]{}
\Text(19.515,10.9162)[br]{}
\Text(20.,-10.)[tr]{}
{#5}
\end{picture}
}

\newcommand{\diagII}[5]{
\begin{picture}(120.,104.)(-28.,-52.)
\Line(0.,0.)(28.,14.)
\Line(28.,14.)(56.,28.)
\Line(56.,28.)(80.,40.)
\Line(28.,-14.)(0.,0.)
\Line(56.,-28.)(28.,-14.)
\Line(80.,-40.)(56.,-28.)
\Photon(56.,28.)(28.,-14.){-2.4}{3}
\Photon(56.,-28.)(28.,14.){-2.4}{3}
\Vertex(56.,28.){2}
\Vertex(56.,-28.){2}
\Vertex(28.,14.){2}
\Vertex(28.,-14.){2}
%external lines
\Text(84.,42.)[l]{#1}
\Text(84.,-42.)[l]{#2}
%gauge bosons
\Text(48.3499,8.07117)[l]{#3}
\Text(48.3499,-9.07117)[l]{#4}
%inflowing lines
\Text(41.5019,21.9679)[br]{}
\Text(42.,-21.)[tr]{}
\Text(19.515,10.9162)[br]{}
\Text(20.,-10.)[tr]{}
{#5}
\end{picture}
}

\newcommand{\diagIII}[6]{
\begin{picture}(120.,104.)(-28.,-52.)
\Line(0.,0.)(24.,12.)
\Line(24.,12.)(56.,28.)
\Line(56.,28.)(80.,40.)
\Line(49.1935,-24.5967)(0.,0.)
\Line(80.,-40.)(49.1935,-24.5967)
\Photon(56.,28.)(49.1935,0.){2.4}{2}
\Photon(24.,12.)(49.1935,0.){-2.4}{2}
\Photon(49.1935,0.)(49.1935,-24.5967){2.4}{2}
\Vertex(56.,28.){2}
\Vertex(24.,12.){2}
\Vertex(49.1935,0.){2}
\Vertex(49.1935,-24.5967){2}
%external lines
\Text(84.,42.)[l]{#1}
\Text(84.,-42.)[l]{#2}
%gauge bosons
\Text(56.5825,13.3573)[l]{#3}
\Text(56.5825,-13.3573)[l]{#4}
\Text(35.307,5.78034)[tr]{#5}
%inflowing lines
\Text(41.5019,21.9679)[br]{}
\Text(32.,-16.)[tr]{}
\Text(19.515,10.9162)[br]{}
{#6}
\end{picture}
}

\newcommand{\diagIIIy}[6]{
\begin{picture}(120.,104.)(-28.,-52.)
\Line(0.,0.)(24.,12.)
\Line(24.,12.)(56.,28.)
\Line(56.,28.)(80.,40.)
\Line(49.1935,-24.5967)(0.,0.)
\Line(80.,-40.)(49.1935,-24.5967)
\DashLine(56.,28.)(49.1935,0.){3}%{2}
\DashLine(24.,12.)(49.1935,0.){3}%{2}
\Photon(49.1935,0.)(49.1935,-24.5967){2.4}{2}
\Vertex(56.,28.){2}
\Vertex(24.,12.){2}
\Vertex(49.1935,0.){2}
\Vertex(49.1935,-24.5967){2}
%external lines
\Text(84.,42.)[l]{#1}
\Text(84.,-42.)[l]{#2}
%gauge bosons
\Text(56.5825,13.3573)[l]{#3}
\Text(56.5825,-13.3573)[l]{#4}
\Text(35.307,5.78034)[tr]{#5}
%inflowing lines
\Text(41.5019,21.9679)[br]{}
\Text(32.,-16.)[tr]{}
\Text(19.515,10.9162)[br]{}
{#6}
\end{picture}
}

\newcommand{\diagIIIb}[6]{
\begin{picture}(120.,104.)(-28.,-52.)
\Line(0.,0.)(24.,-12.)
\Line(24.,-12.)(56.,-28.)
\Line(56.,-28.)(80.,-40.)
\Line(49.1935,24.5967)(0.,0.)
\Line(80.,40.)(49.1935,24.5967)
\Photon(56.,-28.)(49.1935,0.){2.4}{2}
\Photon(24.,-12.)(49.1935,0.){-2.4}{2}
\Photon(49.1935,0.)(49.1935,24.5967){2.4}{2}
\Vertex(56.,-28.){2}
\Vertex(24.,-12.){2}
\Vertex(49.1935,0.){2}
\Vertex(49.1935,24.5967){2}
%external lines
\Text(84.,-42.)[l]{#1}
\Text(84.,42.)[l]{#2}
%gauge bosons
\Text(56.5825,-13.3573)[l]{#3}
\Text(56.5825,13.3573)[l]{#4}
\Text(35.307,5.78034)[tr]{#5}
%inflowing lines
\Text(41.5019,-21.9679)[br]{}
\Text(32.,16.)[tr]{}
\Text(19.515,-10.9162)[br]{}
{#6}
\end{picture}
}

\newcommand{\diagV}[5]{
\begin{picture}(120.,104.)(-28.,-52.)
\Line(0.,0.)(16.,8.)
\Line(16.,8.)(44.,22.)
\Line(44.,22.)(64.,32.)
\Line(64.,32.)(80.,40.)
\Line(64.,-32.)(0.,0.)
\Line(80.,-40.)(64.,-32.)
\Photon(64.,32.)(64.,-32.){2.4}{3.5}
\PhotonArc(30.,15.)(15.6525,26.5651,206.565){2}{4}
\Vertex(64.,-32.){2}
\Vertex(16.,8.){2}
\Vertex(44.,22.){2}
\Vertex(64.,32.){2}
%external lines
\Text(84.,42.)[l]{#1}
\Text(84.,-42.)[l]{#2}
%gauge bosons
\Text(71.5542,0.)[l]{#3}
\Text(23.8885,35.1332)[br]{#4}
%inflowing lines
\Text(41.5019,21.9679)[br]{}
\Text(42.,-21.)[tr]{}
\Text(19.515,10.9162)[br]{}
\Text(20.,-10.)[tr]{}
{#5}
\end{picture}
}

\newcommand{\diagVII}[5]{
\begin{picture}(120.,104.)(-28.,-52.)
\Line(0.,0.)(24.,12.)
\Line(24.,12.)(40.,20.)
\Line(40.,20.)(56.,28.)
\Line(56.,28.)(80.,40.)
\Line(40.,-20.)(0.,0.)
\Line(80.,-40.)(40.,-20.)
\Photon(40.,20.)(40.,-20.){2.4}{3.5}
\PhotonArc(40.,20.)(17.8885,26.5651,206.565){2}{4}
\Vertex(40.,-20.){2}
\Vertex(24.,12.){2}
\Vertex(40.,20.){2}
\Vertex(56.,28.){2}
%external lines
\Text(84.,42.)[l]{#1}
\Text(84.,-42.)[l]{#2}
%gauge bosons
\Text(49.1935,0.)[l]{#3}
\Text(32.1994,42.9325)[br]{#4}
%inflowing lines
\Text(41.5019,21.9679)[br]{}
\Text(42.,-21.)[tr]{}
\Text(19.515,10.9162)[br]{}
\Text(20.,-10.)[tr]{}
{#5}
\end{picture}
}

\newcommand{\diagVy}[5]{
\begin{picture}(120.,104.)(-28.,-52.)
\Line(0.,0.)(16.,8.)
\Line(16.,8.)(44.,22.)
\Line(44.,22.)(64.,32.)
\Line(64.,32.)(80.,40.)
\Line(64.,-32.)(0.,0.)
\Line(80.,-40.)(64.,-32.)
\Photon(64.,32.)(64.,-32.){2.4}{3.5}
\DashCArc(30.,15.)(15.6525,26.5651,206.565){3}%{4}
\Vertex(64.,-32.){2}
\Vertex(16.,8.){2}
\Vertex(44.,22.){2}
\Vertex(64.,32.){2}
%external lines
\Text(84.,42.)[l]{#1}
\Text(84.,-42.)[l]{#2}
%gauge bosons
\Text(71.5542,0.)[l]{#3}
\Text(23.8885,35.1332)[br]{#4}
%inflowing lines
\Text(41.5019,21.9679)[br]{}
\Text(42.,-21.)[tr]{}
\Text(19.515,10.9162)[br]{}
\Text(20.,-10.)[tr]{}
{#5}
\end{picture}
}

\newcommand{\diagIX}[7]{
\begin{picture}(120.,104.)(-28.,-52.)
\Line(0.,0.)(56.,28.)
\Line(56.,28.)(80.,40.)
\Line(56.,-28.)(0.,0.)
\Line(80.,-40.)(56.,-28.)
\Photon(56.,28.)(56.,10.5064){2.4}{2.5}
\Photon(56.,-10.5064)(56.,-28.){2.4}{2.5}
\ArrowArc(56.9771,0.)(11.7466,-90.,90.)
\ArrowArc(56.9771,0.)(11.7466,90.,270.)
\Vertex(56.,-28.){2}
\Vertex(56.,28.){2}
\Vertex(56.,10.5064){2}
\Vertex(56.,-10.5064){2}
%external lines
\Text(84.,42.)[l]{#1}
\Text(84.,-42.)[l]{#2}
%gauge bosons
\Text(64.4276,21.6767)[lu]{#3}
\Text(63.3537,-24.6395)[lb]{#4}
\Text(74.2375,0.)[l]{#5}
\Text(41.1437,0.)[r]{#6}
%inflowing lines
\Text(41.5019,21.9679)[br]{}
\Text(42.,-21.)[tr]{}
\Text(19.515,10.9162)[br]{}
\Text(20.,-10.)[tr]{}
{#7}
\end{picture}
}

\newcommand{\diagVIIy}[5]{
\begin{picture}(120.,104.)(-28.,-52.)
\Line(0.,0.)(24.,12.)
\Line(24.,12.)(40.,20.)
\Line(40.,20.)(56.,28.)
\Line(56.,28.)(80.,40.)
\Line(40.,-20.)(0.,0.)
\Line(80.,-40.)(40.,-20.)
\Photon(40.,20.)(40.,-20.){2.4}{3.5}
\DashCArc(40.,20.)(17.8885,26.5651,206.565){3}%{4}
\Vertex(40.,-20.){2}
\Vertex(24.,12.){2}
\Vertex(40.,20.){2}
\Vertex(56.,28.){2}
%external lines
\Text(84.,42.)[l]{#1}
\Text(84.,-42.)[l]{#2}
%gauge bosons
\Text(49.1935,0.)[l]{#3}
\Text(32.1994,42.9325)[br]{#4}
%inflowing lines
\Text(41.5019,21.9679)[br]{}
\Text(42.,-21.)[tr]{}
\Text(19.515,10.9162)[br]{}
\Text(20.,-10.)[tr]{}
{#5}
\end{picture}
}

\newcommand{\diagself}[3]{
\begin{picture}(120.,104.)(-28.,-52.)
\Line(0.,0.)(56.,28.)
\Line(56.,28.)(80.,40.)
\Line(56.,-28.)(0.,0.)
\Line(80.,-40.)(56.,-28.)
\Photon(56.,28.)(56.,9){2.4}{2.5}
\Photon(56.,-9)(56.,-28.){2.4}{2.5}
\Vertex(56.,-28.){2}
\Vertex(56.,28.){2}
%external lines
\Text(84.,42.)[l]{$\leg{i}$}
\Text(84.,-42.)[l]{$\leg{j}$}
%gauge bosons
\Text(64.4276,21.6767)[lu]{#1}
\Text(63.3537,-24.6395)[lb]{#2}
%inflowing lines
\Text(41.5019,21.9679)[br]{}
\Text(42.,-21.)[tr]{}
\Text(19.515,10.9162)[br]{}
\Text(20.,-10.)[tr]{}
\GCirc(56,0.){9}{0.9}
{#3}
\end{picture}
}

\newcommand{\diagX}[1]{
\begin{picture}(120.,104.)(-28.,-52.)
\Line(0.,0.)(56.,28.)
\Line(56.,28.)(80.,40.)
\Line(56.,-28.)(0.,0.)
\Line(80.,-40.)(56.,-28.)
\Photon(56.,28.)(56.,10.5064){2.4}{2.5}
\Photon(56.,-10.5064)(56.,-28.){2.4}{2.5}
\PhotonArc(56.9771,0.)(11.7466,-101.459,258.541){-2}{8}
\Vertex(56.,-28.){2}
\Vertex(56.,28.){2}
\Vertex(56.,10.5064){2}
\Vertex(56.,-10.5064){2}
%external lines
\Text(84.,42.)[l]{$\leg{i}$}
\Text(84.,-42.)[l]{$\leg{j}$}
%gauge bosons
\Text(64.4276,21.6767)[lu]{$\scriptscriptstyle{V_1}$}
\Text(76.0263,0.)[l]{$\scriptscriptstyle{V_2}$}
\Text(38.4604,0.)[r]{$\scriptscriptstyle{V_3}$}
\Text(63.3537,-24.6395)[lb]{$\scriptscriptstyle{V_4}$}
%inflowing lines
\Text(41.5019,21.9679)[br]{}
\Text(42.,-21.)[tr]{}
\Text(19.515,10.9162)[br]{}
\Text(20.,-10.)[tr]{}
{#1}
\end{picture}
}

\newcommand{\diagXI}[1]{
\begin{picture}(120.,104.)(-28.,-52.)
\Line(0.,0.)(56.,28.)
\Line(56.,28.)(80.,40.)
\Line(56.,-28.)(0.,0.)
\Line(80.,-40.)(56.,-28.)
\Photon(56.,28.)(56.,10.5064){2.4}{2.5}
\Photon(56.,-10.5064)(56.,-28.){2.4}{2.5}
\DashArrowArcn(56.9771,0.)(11.7466,90.,-90.){1}
\DashArrowArcn(56.9771,0.)(11.7466,270.,90.){1}
\Vertex(56.,-28.){2}
\Vertex(56.,28.){2}
\Vertex(56.,10.5064){2}
\Vertex(56.,-10.5064){2}
%external lines
\Text(84.,42.)[l]{$\leg{i}$}
\Text(84.,-42.)[l]{$\leg{j}$}
%gauge bosons
\Text(64.4276,21.6767)[lu]{$\scriptscriptstyle{V_1}$}
\Text(73.343,1.)[l]{$\scriptstyle{u^{V_2}}$}
\Text(41.1437,1.)[r]{$\scriptstyle{u^{V_3}}$}
\Text(63.3537,-24.6395)[lb]{$\scriptscriptstyle{V_4}$}
%inflowing lines
\Text(41.5019,21.9679)[br]{}
\Text(42.,-21.)[tr]{}
\Text(19.515,10.9162)[br]{}
\Text(20.,-10.)[tr]{}
{#1}
\end{picture}
}

\newcommand{\diagXII}[1]{
\begin{picture}(120.,104.)(-28.,-52.)
\Line(0.,0.)(56.,28.)
\Line(56.,28.)(80.,40.)
\Line(56.,-28.)(0.,0.)
\Line(80.,-40.)(56.,-28.)
\Photon(56.,28.)(56.,10.5064){2.4}{2.5}
\Photon(56.,-10.5064)(56.,-28.){2.4}{2.5}
\DashCArc(56.9771,0.)(11.7466,-90.,90.){3}
\DashCArc(56.9771,0.)(11.7466,90.,270.){3}
\Vertex(56.,-28.){2}
\Vertex(56.,28.){2}
\Vertex(56.,10.5064){2}
\Vertex(56.,-10.5064){2}
%external lines
\Text(84.,42.)[l]{$\leg{i}$}
\Text(84.,-42.)[l]{$\leg{j}$}
%gauge bosons
\Text(64.4276,21.6767)[lu]{$\scriptscriptstyle{V_1}$}
\Text(71.5542,0.)[l]{$\scriptscriptstyle{\Phi_{i_2}}$}
\Text(42.9325,0.)[r]{$\scriptscriptstyle{\Phi_{i_3}}$}
\Text(63.3537,-24.6395)[lb]{$\scriptscriptstyle{V_4}$}
%inflowing lines
\Text(41.5019,21.9679)[br]{}
\Text(42.,-21.)[tr]{}
\Text(19.515,10.9162)[br]{}
\Text(20.,-10.)[tr]{}
{#1}
\end{picture}
}

\newcommand{\diagXV}[1]{
\begin{picture}(120.,104.)(-28.,-52.)
\Line(0.,0.)(56.,28.)
\Line(56.,28.)(80.,40.)
\Line(56.,-28.)(0.,0.)
\Line(80.,-40.)(56.,-28.)
\Photon(56.,28.)(56.,10.5064){-2.4}{2.5}
\Photon(56.,-10.5064)(56.,-28.){-2.4}{2.5}
\PhotonArc(56.9771,0.)(11.7466,90.,270.){-2}{3.5}
\DashCArc(56.9771,0.)(11.7466,-90.,90.){3}
\Vertex(56.,-28.){2}
\Vertex(56.,28.){2}
\Vertex(56.,10.5064){2}
\Vertex(56.,-10.5064){2}
%external lines
\Text(84.,42.)[l]{$\leg{i}$}
\Text(84.,-42.)[l]{$\leg{j}$}
%gauge bosons
\Text(64.4276,21.6767)[lu]{$\scriptscriptstyle{V_1}$}
\Text(76.0263,0.)[l]{$\scriptscriptstyle{\Phi_{i_2}}$}
\Text(38.4604,0.)[r]{$\scriptscriptstyle{V_3}$}
\Text(63.3537,-24.6395)[lb]{$\scriptscriptstyle{V_4}$}
%inflowing lines
\Text(41.5019,21.9679)[br]{}
\Text(42.,-21.)[tr]{}
\Text(19.515,10.9162)[br]{}
\Text(20.,-10.)[tr]{}
{#1}
\end{picture}
}

\newcommand{\diagXVI}[1]{
\begin{picture}(120.,104.)(-28.,-52.)
\Line(0.,0.)(48.,24.)
\Line(48.,24.)(80.,40.)
\Line(48.,-24.)(0.,0.)
\Line(80.,-40.)(48.,-24.)
\Photon(48.,24.)(58.1378,0.){-2.4}{2.5}
\Photon(48.,-24.)(58.1378,0.){2.4}{2.5}
\PhotonArc(67.082,0.)(8.94427,-177.135,182.865){-2}{7}
\Vertex(48.,-24.){2}
\Vertex(48.,24.){2}
\Vertex(58.1378,0.){2}
%external lines
\Text(84.,42.)[l]{$\leg{i}$}
\Text(84.,-42.)[l]{$\leg{j}$}
%gauge bosons
\Text(55.9503,18.8245)[lu]{$\scriptscriptstyle{V_1}$}
\Text(80.4984,0.)[l]{$\scriptscriptstyle{V_2}$}
\Text(55.0177,-21.3975)[lb]{$\scriptscriptstyle{V_3}$}
%inflowing lines
\Text(41.5019,21.9679)[br]{}
\Text(42.,-21.)[tr]{}
\Text(19.515,10.9162)[br]{}
\Text(20.,-10.)[tr]{}
{#1}
\end{picture}
}

\newcommand{\diagXVII}[1]{
\begin{picture}(120.,104.)(-28.,-52.)
\Line(0.,0.)(48.,24.)
\Line(48.,24.)(80.,40.)
\Line(48.,-24.)(0.,0.)
\Line(80.,-40.)(48.,-24.)
\Photon(48.,24.)(58.1378,0.){-2.4}{2.5}
\Photon(48.,-24.)(58.1378,0.){2.4}{2.5}
\DashCArc(67.082,0.)(8.94427,-177.135,182.865){3}
\Vertex(48.,-24.){2}
\Vertex(48.,24.){2}
\Vertex(58.1378,0.){2}
%external lines
\Text(84.,42.)[l]{$\leg{i}$}
\Text(84.,-42.)[l]{$\leg{j}$}
%gauge bosons
\Text(55.9503,18.8245)[lu]{$\scriptscriptstyle{V_1}$}
\Text(80.4984,0.)[l]{$\scriptscriptstyle{\Phi_{i_2}}$}
\Text(55.0177,-21.3975)[lb]{$\scriptscriptstyle{V_3}$}
%inflowing lines
\Text(41.5019,21.9679)[br]{}
\Text(42.,-21.)[tr]{}
\Text(19.515,10.9162)[br]{}
\Text(20.,-10.)[tr]{}
{#1}
\end{picture}
}

\newcommand{\diagAZ}[1]{
\begin{picture}(120.,104.)(-28.,-52.)
\Line(0.,0.)(56.,28.)
\Line(56.,28.)(80.,40.)
\Line(56.,-28.)(0.,0.)
\Line(80.,-40.)(56.,-28.)
\Photon(56.,28.)(56.,10.5064){-2.4}{2.5}
\Photon(56.,-10.5064)(56.,-28.){-2.4}{2.5}
\GCirc(56.,0.){11}{0.8}
\Vertex(56.,-28.){2}
\Vertex(56.,28.){2}
%external lines
\Text(84.,42.)[l]{$\leg{i}$}
\Text(84.,-42.)[l]{$\leg{j}$}
%gauge bosons
\Text(64.4276,21.6767)[lu]{$\scriptscriptstyle{A}$}
\Text(76.0263,0.)[l]{}
\Text(38.4604,0.)[r]{}
\Text(63.9079,-23.1643)[lb]{$\scriptscriptstyle{Z}$}
%inflowing lines
\Text(41.5019,21.9679)[br]{}
\Text(42.,-21.)[tr]{}
\Text(19.515,10.9162)[br]{}
\Text(20.,-10.)[tr]{}
{#1}
\end{picture}
}

\newcommand{\diagXX}[6]{
\begin{picture}(120.,104.)(-28.,-52.)
\Line(0.,0.)(65.,32.5)
\Line(0.,0.)(65.,-32.5)
\Line(0.,0.)(72.6722,0.)
\Photon(55.25,27.625)(55.25,0.){1.95}{2.5}
\Photon(37.05,-18.525)(37.05,0.){-1.95}{2}
\Vertex(55.25,27.625){2}
\Vertex(37.05,-18.525){2}
\Vertex(55.25,0.){2}
\Vertex(37.05,0.){2}
%external lines
\Text(68.25,34.125)[l]{#1}
\Text(76.3058,0.)[l]{#2}
\Text(68.25,-34.125)[l]{#3}
%gauge bosons
\Text(60.4318,13.7948)[l]{#4}
\Text(42.4369,-13.018)[l]{#5}
%inflowing lines
\Text(46.5102,0.)[b]{}
\Text(32.1146,16.999)[br]{}
\Text(22.75,-11.375)[tr]{}
\Text(25.4353,0.)[b]{}
{#6}
\end{picture}
}

\newcommand{\diagXXd}[6]{
\begin{picture}(120.,104.)(-28.,-52.)
\Line(0.,0.)(65.,32.5)
\Line(0.,0.)(65.,-32.5)
\Line(0.,0.)(72.6722,0.)
\PhotonArc(30.,15.)(15.6525,26.5651,206.565){2}{4}
\Vertex(16.,8.){2}
\Vertex(44.,22.){2}
\Photon(50,-25)(50,0.){-1.95}{2}
\Vertex(50,-25){2}
\Vertex(50,0.){2}
%external lines
\Text(68.25,34.125)[l]{#1}
\Text(76.3058,0.)[l]{#2}
\Text(68.25,-34.125)[l]{#3}
%gauge bosons
\Text(10.4318,30.7948)[b]{#4}
\Text(55,-15)[l]{#5}
%inflowing lines
\Text(46.5102,0.)[b]{}
\Text(32.1146,16.999)[br]{}
\Text(22.75,-11.375)[tr]{}
\Text(25.4353,0.)[b]{}
{#6}
\end{picture}
}

\newcommand{\diagXXc}[6]{
\begin{picture}(120.,104.)(-28.,-52.)
\Line(0.,0.)(65.,32.5)
\Line(0.,0.)(65.,-32.5)
\Line(0.,0.)(72.6722,0.)
\Photon(37.05,18.525)(37.05,0.){1.95}{2}
\Photon(55.25,-27.625)(55.25,0.){-1.95}{2.5}
\Vertex(37.05,18.525){2}
\Vertex(55.25,-27.625){2}
\Vertex(37.05,0.){2}
\Vertex(55.25,0.){2}
%external lines
\Text(68.25,34.125)[l]{#1}
\Text(76.3058,0.)[l]{#2}
\Text(68.25,-34.125)[l]{#3}
%gauge bosons
\Text(60.1189,-14.1922)[l]{#4}
\Text(42.4369,10.018)[l]{#5}
%inflowing lines
\Text(46.5102,0.)[t]{}
\Text(32.1146,16.999)[br]{}
\Text(22.75,-11.375)[tr]{}
\Text(25.4353,0.)[t]{}
{#6}
\end{picture}
}

\newcommand{\diagXXI}[7]{
\begin{picture}(120.,104.)(-28.,-52.)
\Line(0.,0.)(65.,32.5)
\Line(0.,0.)(65.,-32.5)
\Line(0.,0.)(72.36,-6.72921)
\Photon(45.5,22.75)(50.3792,7.05301){1.95}{2}
\Photon(39.,-19.5)(50.3792,7.05301){-1.95}{3}
\Photon(62.9532,-5.85441)(50.3792,7.05301){1.95}{2}
\Vertex(45.5,22.75){2}
\Vertex(62.9532,-5.85441){2}
\Vertex(39.,-19.5){2}
\Vertex(50.3792,7.05301){2}
%external lines
\Text(68.25,34.125)[l]{#1}
\Text(75.978,-7.06567)[l]{#2}
\Text(68.25,-34.125)[l]{#3}
%gauge bosons
\Text(52.0477,16.1784)[l]{#4}
\Text(61.6758,3.43505)[l]{#5}
\Text(46.4209,-13.9719)[l]{#6}
%inflowing lines
\Text(26.,-13.)[tr]{}
\Text(28.9031,15.2991)[br]{}
\Text(29.0689,0.)[b]{}
{#7}
\end{picture}
}

\newcommand{\diagXXII}[7]{
\begin{picture}(120.,104.)(-28.,-52.)
\Line(0.,0.)(65.,32.5)
\Line(0.,0.)(65.,-32.5)
\Line(0.,0.)(72.36,6.72921)
\Line(0.,0.)(72.36,-6.72921)
\Photon(52.,26.)(57.888,5.38336){1.95}{2}
\Photon(52.,-26.)(57.888,-5.38336){-1.95}{2}
\Vertex(52.,26.){2}
\Vertex(52.,-26.){2}
\Vertex(57.888,5.38336){2}
\Vertex(57.888,-5.38336){2}
%external lines
\Text(68.25,34.125)[l]{#1}
\Text(75.978,7.06567)[l]{#2}
\Text(75.978,-7.06567)[l]{#3}
\Text(68.25,-34.125)[l]{#4}
%gauge bosons
\Text(59.3965,16.9633)[l]{#5}
\Text(59.3965,-16.9633)[l]{#6}
{#7}
\end{picture}
}

\newcommand{\diagonenf}[3]{
%\begin{picture}(120.,104.)(-28.,-52.)
\begin{picture}(120.,76.)(-28.,-38.)
\Line(0.,0.)(80,0)
\PhotonArc(25.,-8.)(26.2488,17.7447,162.255){2}{5}
\Vertex(50.,0.){2}
\Text(84.,0)[l]{#1}
%gauge bosons
\Text(23,29)[l]{#2}
%inflowing lines
\Text(27.668,14.6453)[br]{}
\Text(28.,-14.)[tr]{}
{#3}
\end{picture}
}

\newcommand{\diagoneselfnf}[3]{
%\begin{picture}(120.,104.)(-28.,-52.)
\begin{picture}(120.,76.)(-28.,-38.)
\Line(0.,0.)(80,0)
\PhotonArc(40.,0)(20,0,180){2}{5}
\Vertex(60.,0.){2}
\Vertex(20.,0.){2}
\Text(84.,0)[l]{#1}
\Text(40,29.)[b]{#2}
\Text(27.668,14.6453)[br]{}
\Text(28.,-14.)[tr]{}
{#3}
\end{picture}
}

\newcommand{\diagonefact}[4]{
\begin{picture}(120.,104.)(-28.,-52.)
\Line(0.,0.)(80,0)
\Line(0.,0.)(50,50)
\PhotonArc(0,0)(48,0,45){2}{4}
\Vertex(48.,0.){2}
\Vertex(33.941,33.941){2}
\Text(84.,0)[l]{#1}
\Text(53.,53.)[l]{#2}
%gauge bosons
\Text(50,22.)[l]{#3}
%inflowing lines
\Text(27.668,14.6453)[br]{}
\Text(28.,-14.)[tr]{}
{#4}
\end{picture}
}

\newcommand{\diagInf}[6]{
\begin{picture}(120.,104.)(-28.,-52.)
\Line(0.,0.)(28.,14.)
\Line(28.,14.)(56.,28.)
\Line(56.,28.)(80.,40.)
\Line(28.,-14.)(0.,0.)
\Line(56.,-28.)(28.,-14.)
\Line(80.,-40.)(56.,-28.)
\Photon(56.,28.)(56.,0.){2.4}{2}
\Photon(56.,0.)(56.,-28.){2.4}{2}
\Photon(56.,0.)(0,0){2.4}{4}
\Vertex(56.,28.){2}
\Vertex(56.,0){2}
\Vertex(56.,-28.){2}
%external lines
\Text(84.,42.)[l]{#1}
\Text(84.,-42.)[l]{#2}
%gauge bosons
\Text(62.6099,14.)[l]{#3}
\Text(62.6099,-14.)[l]{#4}
\Text(35.305,-5.)[t]{#5}
%inflowing lines
\Text(41.5019,21.9679)[br]{}
\Text(42.,-21.)[tr]{}
\Text(19.515,10.9162)[br]{}
\Text(20.,-10.)[tr]{}
{#6}
\end{picture}
}
\newcommand{\diagIInf}[5]{
\begin{picture}(120.,104.)(-28.,-52.)
\Line(0.,0.)(16.,8.)
\Line(16.,8.)(44.,22.)
\Line(44.,22.)(64.,32.)
\Line(64.,32.)(80.,40.)
\Line(64.,-32.)(0.,0.)
\Line(80.,-40.)(64.,-32.)
\Photon(64.,32.)(64.,-32.){2.4}{3.5}
\PhotonArc(17.9434,8.83368)(20.,26.2114,206.211){2}{5} 
\Vertex(35.8868,17.6674){2}
\Vertex(64.,-32.){2}
\Vertex(64.,32.){2}
%external lines
\Text(84.,42.)[l]{#1}
\Text(84.,-42.)[l]{#2}
%gauge bosons
\Text(71.5542,0.)[l]{#3}
\Text(20.8885,35.1332)[br]{#4}
%inflowing lines
\Text(41.5019,21.9679)[br]{}
\Text(42.,-21.)[tr]{}
\Text(19.515,10.9162)[br]{}
\Text(20.,-10.)[tr]{}
{#5}
\end{picture}
}

\newcommand{\diagIIInf}[5]{
\begin{picture}(120.,104.)(-28.,-52.)
\Line(0.,0.)(24.,12.)
\Line(24.,12.)(40.,20.)
\Line(40.,20.)(56.,28.)
\Line(56.,28.)(80.,40.)
\Line(40.,-20.)(0.,0.)
\Line(80.,-40.)(40.,-20.)
\Photon(40.,20.)(40.,-20.){2.4}{3.5}
\PhotonArc(44.6515,-11.4562)(46.0977,66.8127,165.61){2}{5}
\Vertex(62.8019,30.9179){2}
\Vertex(40.,-20.){2}
\Vertex(40.,20.){2}
%external lines
\Text(84.,42.)[l]{#1}
\Text(84.,-42.)[l]{#2}
%gauge bosons
\Text(49.1935,0.)[l]{#3}
\Text(30.1994,40.9325)[br]{#4}
%inflowing lines
\Text(41.5019,21.9679)[br]{}
\Text(42.,-21.)[tr]{}
\Text(19.515,10.9162)[br]{}
\Text(20.,-10.)[tr]{}
{#5}
\end{picture}
}

\newcommand{\diagIVnf}[6]{
\begin{picture}(120.,104.)(-28.,-52.)
\Line(0.,0.)(65.,32.5)
\Line(0.,0.)(65.,-32.5)
\Line(0.,0.)(72.6722,0.)
\PhotonArc(23.7198,2.76053)(23.8799,45.7845,186.638){2}{5}
\Vertex(40.3727,19.8758){2}
\Photon(50,-25)(50,0.){-1.95}{2}
\Vertex(50,-25){2}
\Vertex(50,0.){2}
%external lines
\Text(68.25,34.125)[l]{#1}
\Text(76.3058,0.)[l]{#2}
\Text(68.25,-34.125)[l]{#3}
%gauge bosons
\Text(10.4318,30.7948)[b]{#4}
\Text(55,-15)[l]{#5}
%inflowing lines
\Text(46.5102,0.)[b]{}
\Text(32.1146,16.999)[br]{}
\Text(22.75,-11.375)[tr]{}
\Text(25.4353,0.)[b]{}
{#6}
\end{picture}
}

\newcommand{\diagIwi}[3]{
%\begin{picture}(120.,104.)(-28.,-52.)
\begin{picture}(120.,64.)(-28.,-32.)
\Line(0.,0.)(50,0)
\PhotonArc(50.,-50.)(70.71,90.,135.){2}{5}
\Text(54.,0)[l]{#1}
%gauge bosons
\Text(54,22.)[l]{#2}
%inflowing lines
\Text(27.668,14.6453)[br]{}
\Text(28.,-14.)[tr]{}
{#3}
\end{picture}
}
\newcommand{\diagIIwi}[3]{
%\begin{picture}(120.,104.)(-28.,-52.)
\begin{picture}(120.,64.)(-28.,-32.)
\Line(0.,0.)(50,0)
\Photon(25.,0)(50,20){-2}{3}
\Vertex(25,0.){2}
\Text(54.,0)[l]{#1}
\Text(54,22.)[l]{#2}
%inflowing lines
\Text(27.668,14.6453)[br]{}
\Text(28.,-14.)[tr]{}
{#3}
\end{picture}
}

\newcommand{\diagIIIwi}[2]{
\begin{picture}(120.,104.)(-28.,-52.)
\Line(0.,0.)(50,0)
\Text(54.,0)[l]{#1}
%gauge bosons
%inflowing lines
\Text(27.668,14.6453)[br]{}
\Text(28.,-14.)[tr]{}
{#2}
\end{picture}
}

\newcommand{\diagIVwi}[4]{
\begin{picture}(120.,104.)(-28.,-52.)
\Line(0.,0.)(70,0)
\Line(0.,0.)(50,50)
\Photon(25,25)(70,25){2}{5}
\Vertex(25,25){2}
\Text(74.,0)[l]{#1}
\Text(53.,53.)[l]{#2}
%gauge bosons
\Text(75,25)[l]{#3}
%inflowing lines
\Text(27.668,14.6453)[br]{}
\Text(28.,-14.)[tr]{}
{#4}
\end{picture}
}

\newcommand{\subloopI}[4]{
\begin{picture}(120.,104.)(-28.,-52.)
\Line(0.,0.)(16.,8.)
\Line(16.,8.)(44.,22.)
\Line(44.,22.)(64.,32.)
\Line(64.,32.)(80.,40.)
\Line(64.,-32.)(0.,0.)
\Line(80.,-40.)(64.,-32.)
\Photon(64.,32.)(64.,-32.){2.4}{3.5}
\GCirc(35.,17.5){10}{0.8}
\Vertex(64.,-32.){2}
\Vertex(64.,32.){2}
%external lines
\Text(84.,42.)[l]{#1}
\Text(84.,-42.)[l]{#2}
%gauge bosons
\Text(71.5542,0.)[l]{#3}
%inflowing lines
\Text(41.5019,21.9679)[br]{}
\Text(42.,-21.)[tr]{}
\Text(19.515,10.9162)[br]{}
\Text(20.,-10.)[tr]{}
{#4}
\end{picture}
}

\newcommand{\subloopII}[4]{
\begin{picture}(120.,104.)(-28.,-52.)
\Line(0.,0.)(24.,12.)
\Line(24.,12.)(40.,20.)
\Line(40.,20.)(56.,28.)
\Line(56.,28.)(80.,40.)
\Line(40.,-20.)(0.,0.)
\Line(80.,-40.)(40.,-20.)
\Photon(40.,20.)(40.,-20.){2.4}{3.5}
\GCirc(40.,20){10}{0.8}
\Vertex(40.,-20.){2}
%external lines
\Text(84.,42.)[l]{#1}
\Text(84.,-42.)[l]{#2}
%gauge bosons
\Text(49.1935,0.)[l]{#3}
%inflowing lines
\Text(41.5019,21.9679)[br]{}
\Text(42.,-21.)[tr]{}
\Text(19.515,10.9162)[br]{}
\Text(20.,-10.)[tr]{}
{#4}
\end{picture}
}

\newcommand{\diagIcollin}[4]{
%\begin{picture}(100.,104.)(-15.,-52.)
\begin{picture}(100.,88.)(-15.,-44.)
\Line(0.,0.)(90,0)
\Text(94.,0)[l]{#1}
\Photon(0,0)(0,-30){2}{4}
\Text(0,-34)[t]{#2}
\Photon(55,0)(55,-30){2}{4}
\Vertex(55,0){2}
\Text(55,-34)[t]{#3}
{#4}
\end{picture}
}

\newcommand{\diagIIcollin}[6]{
%\begin{picture}(175.,104.)(-5.,-52.)
\begin{picture}(175.,88.)(-5.,-44.)
\Photon(0,0)(-15,-30){2}{4}
\Text(0,-34)[t]{#1}
\Photon(0,0)(15,-30){2}{4}
\Text(15,-34)[t]{#2}
\Line(0.,0.)(80,0)
\Text(90,0)[l]{$\dots$}
\Line(120.,0.)(170,0)
\Text(174.,0)[l]{#3}
\Photon(65,0)(65,-30){2}{4}
\Vertex(65,0){2}
\Text(65,-34)[t]{#4}
\Photon(135,0)(135,-30){2}{4}
\Vertex(135,0){2}
\Text(135,-34)[t]{#5}
{#6}
\end{picture}
}

%%%%%%%%%%%%%%%%%%%%%%  abbreviations for results   %%%%%%%%%%%%%%%%%%%%%%
\newcommand{\Frac}{\frac}
\newcommand{\Epsinv}[1]{\veps^{- #1}}
\newcommand{\Eps}[1]{\veps^{#1}}
\newcommand{\Right}{\right}
\newcommand{\EG}{\gamma_{\mathrm{E}}}

%%%%%%%%%%%%%%%%%%%%% %%%%%%%%%%%%%%%%%%%%%%%%%%%%%%%%%%%%%%%%%

\section{Introduction}
\label{se:intro}%\refse{intro}
%%%%%%%%%%%%%%%%%%%%%%%%%%%%%%%%%%%%%%%%%%%%%%%%%%
% EW corrections at TeV scale
%%%%%%%%%%%%%%%%%%%%%%%%%%%%%%%%%%%%%%%%%%%%%%%%%%
At TeV colliders, the electroweak corrections
are strongly enhanced by logarithms
of the form $\alpha^l\ln^{j}{\left({Q^2}/{M^2}\right)}$, with  
$M=\MW\sim\MZ$ and $j \le 2l$.
These logarithmic corrections affect every reaction that involves
electroweakly interacting particles and is characterized by scattering
energies $Q^2\gg M^2$.
For $Q\sim 1\TeV$ their impact can reach tens of per cent at one loop and several per cent
at two loops (see for instance \citeres{Kuhn:2005az,Kuhn:2007cv} and
references therein).
Their origin is twofold.  The renormalization of UV singularities at
the scale $\muR\lsim M$ gives rise to terms of the form
$\ln(Q^2/\muR^2)$.  In addition, the interactions of the initial- and
final-state particles with soft and/or collinear gauge bosons give
rise to mass-singular logarithms, \ie $\ln(Q^2/M^2)$ terms that are
formally singular in the limit $M\to 0$.
Since they originate from UV, soft and collinear singularities,
electroweak logarithmic corrections have universal properties that
can be studied in a process-independent way and exhibit interesting
analogies with QCD.
At one loop, the leading logarithms (LLs) and next-to-leading
logarithms (NLLs) factorize and are described by a general formula
that applies to arbitrary Standard-Model processes
\cite{Denner:2001jv,Denner:2001gw}.
The properties of electroweak logarithmic corrections beyond one loop
are investigated with two complementary approaches in the literature:
(i) Evolution equations, which are well known in QED and
QCD, are applied to the electroweak theory in order to obtain
the higher-order terms through a resummation
\cite{Fadin:2000bq,Kuhn:2000nn,Melles:2001ye,Kuhn:2001hz,Kuhn:2007ca}. In this approach the evolution is split into two regimes, where the
electroweak interaction is described in terms of SU(2)$\times$U(1)
and $\mathrm{U}_{\mathrm{em}}(1)$ symmetric gauge theories.%
\footnote{
A new approach based on soft-collinear effective theory has been
developed in \citere{Chiu:2007dg}.}
This splitting is assumed to correctly reproduce 
all relevant implications of electroweak symmetry
breaking in LL, NLL and NNLL approximation.
(ii) This assumption and the resulting resummation prescriptions 
can be checked and improved by means 
of explicit diagrammatic two-loop
calculations based on the (spontaneously broken) electroweak Lagrangian.
So far, all existing diagrammatic results 
are in agreement with the resummation prescriptions,
however up to now only a small subset of logarithms and 
processes has been computed explicitly.
While the LLs
\cite{Beenakker:2001kf} and the
angular-dependent subset of the NLLs \cite{Denner:2003wi}
have been derived for arbitrary processes,
complete diagrammatic
calculations at (or beyond) the NLL level exist only for
matrix elements involving massless external fermions
\cite{Pozzorini:2004rm,Jantzen:2005az,Denner:2006jr}.
In the literature, no explicit NLL calculation exists for reactions
involving massive scattering particles.

In the following we present a recent two-loop NLL calculation \cite{nextpaper}
for $n$-fermion processes $f_1f_2\to f_3\dots f_n$ involving an arbitrary
number of leptons and/or quarks.  While all fermions with $m_f\ll M$
are treated as massless, the top-quark mass effects are taken into
account.  For all invariants
$r_{ij}=(p_i+p_j)^2$
we assume $|r_{ij}|\sim Q^2 \gg M^2$, and 
the differences between the W, Z, H and t mass are 
taken into account.
Soft and collinear singularities from massless virtual photons are
regularized dimensionally and arise as $\veps$-poles in $D=4-2\veps$
dimensions.  
For consistency, the same power counting is applied to 
$\ln(Q^2/M^2)$ and $1/\veps$ singularities. Thus,
in
NLL approximation we include all  $\veps^{-k}\ln^{j-k}(Q^2/M^2)$ terms
with total power  $j=2,1$ at one loop and $j=4,3$ at two loops.
The photonic singularities are factorized
in a gauge-invariant electromagnetic term.  The remaining part of the
corrections -- which is finite, gauge invariant, and does not depend on the
scheme adopted to regularize photonic singularities -- contains only
$\ln(Q^2/M^2)$ terms.  The divergences contained in the
electromagnetic term cancel if real-photon emission is included.

\section{Extraction of soft, collinear and UV
contributions from Feynman diagrams}
Let us consider the two-loop diagrams that produce NLLs within the
't~Hooft--Feynman gauge.  These diagrams are obtained from the LL one-loop
diagrams, \ie diagrams with external-leg exchange of soft-collinear gauge
bosons, via insertions of one-loop sub-diagrams that produce an additional
logarithm.  We start with the diagrams that do not contain Yukawa
interactions, 
\unitlength 0.6pt\SetScale{0.6}
\beqar\label{twoloopdiag2}%\refeq{twoloopdiag2} \!\!
\hspace{-4mm}
\vcenter{\hbox{\diagIInf{$\leg{i}$}{$\leg{j}$}{$\scriptscriptstyle{V_1}$}{$\scriptscriptstyle{V_2}$}{\blob}}}
\hspace{-2mm},
\vcenter{\hbox{\diagIIInf{$\leg{i}$}{$\leg{j}$}{$\scriptscriptstyle{V_1}$}{$\scriptscriptstyle{V_2}$}{\blob}}}
\hspace{-2mm},
\vcenter{\hbox{\diagIVnf{$\leg{j}$}{$\leg{i}$}{$\leg{k}$}{$\scriptscriptstyle{V_2}$}{$\scriptscriptstyle{V_1}$}{\blob}}}
\hspace{-2mm},
\vcenter{\hbox{\diagInf{$\leg{i}$}{$\leg{j}$}{$\scriptscriptstyle{V_1}$}{$\scriptscriptstyle{V_3}$}{$\scriptscriptstyle{V_2}$}{\blob}}}
\hspace{-2mm},
\vcenter{\hbox{\diagself{$\scriptscriptstyle{V_1}$}{$\scriptscriptstyle{V_2}$}{\blob}}}
.
\eeqar
Here $V_i=A,Z,W^\pm$, and the external lines $i,j,k=1,\dots,n$ are massless and/or massive fermions.
In the first four diagrams in \refeq{twoloopdiag2} the gauge boson
$V_2$ can couple to an external line or to an internal propagator.  In
the latter case the only relevant region is the one where $V_2$ is
collinear to an external fermion and, using collinear Ward identities, we
find that this type of contributions cancel
\cite{Denner:2006jr,nextpaper}.  
The remaining diagrams contain 
a tree sub-diagram that corresponds to the 
original process and depends only on the external momenta.%
\footnote{In these tree sub-diagrams, which are denoted with an 'F' in
  \refeq{twoloopnf4}, the loop momenta must be set to zero.  This
  follows from the collinear Ward identities.}
These diagrams are called factorizable, 
since their NLL contribution factorizes into the 
$n$-fermion tree amplitude times a two-loop NLL factor.
For instance, the factorizable diagrams associated with the fourth diagram in 
\refeq{twoloopdiag2} are
 {
\unitlength 0.6pt\SetScale{0.6}
\beqar\label{twoloopnf4}%\refeq{twoloopnf4}
\vcenter{\hbox{\diagIII{$\leg{i}$}{$\leg{j}$}{$\scriptscriptstyle{V_1}$}{$\scriptscriptstyle{V_3}$}{$\scriptscriptstyle{V_2}$}{\factblob}}}
,\quad
\vcenter{\hbox{\diagIIIb{$\leg{j}$}{$\leg{i}$}{$\scriptscriptstyle{V_3}$}{$\scriptscriptstyle{V_1}$}{$\scriptscriptstyle{V_2}$}{\factblob}}}
,\quad
\vcenter{\hbox{\diagXXI{$\leg{k}$}{$\leg{i}$}{$\leg{j}$}{$\scriptscriptstyle{V_2}$}{$\scriptscriptstyle{V_1}$}{$\scriptscriptstyle{V_3}$}{\factblob}}}
.
\eeqar
}%
To extract the NLLs of
soft/collinear origin,
we exploit the fact that
in the soft/collinear limit the coupling of a gauge boson $V$ to the $i$-th
external fermion yields a factor  $- 2 e I^V_i (p_i+q)^\mu$,
where $q$ and $p_i$ are the gauge-boson and fermion momenta, while $I^V_i$ is the gauge-group generator.
For instance, for the last diagram in \refeq{twoloopnf4} we obtain
\beqar\label{twoloopnonfactpartD}%\refeq{twoloopnonfactpartD}
\muD^{4\epsilon}
\int\frac{\rd^D q_1\rd^D q_2}{({2\pi})^{2D}}
\frac{-8 \ri e^3\gw {\teps^{\bar V_1 \bar V_2 \bar V_3}}\;
\left[
g_{\mu_1\mu_2}(q_1-q_2)_{\mu_3}
+\mbox{perm.}
\right]
l_1^{\mu_1}
l_2^{\mu_2}
l_3^{\mu_3}
}{
(q_1^2-M_{V_1}^2)
(q_2^2-M_{V_2}^2)
(q_3^2-M_{V_3}^2)
(l^2_1-m^2_1)
(l^2_2-m^2_2)
(l^2_3-m^2_3)
}
\times
I^{{V}_1}_i
I^{{V}_2}_k
I^{{V}_3}_j
\M_0,
\nn
\eeqar
where 
$q_3=-q_1-q_2$,
$l_1=p_i-q_1$,
$l_2=p_k-q_2$ and
$l_3=p_j-q_3$.
The $n$-fermion tree amplitude $\M_0$ must be regarded as a vector 
carrying SU(2) indices associated with the isospin of the
scattering fermions, and 
the generators $I^{V}_i$
act as SU(2) matrices on the $i$-th fermion.

In addition to soft/collinear contributions, there are also two-loop
NLL terms that arise from UV divergences in the one-loop sub-diagrams.
In this
case, the UV singularities are removed by means of a minimal
subtraction at the scale $\mu^2=Q^2$. As a consequence, only those
UV-divergent sub-diagrams characterized by scales of order $M^2\ll Q^2$
produce NLL terms. In particular, non-factorizable diagrams do not
produce NLL terms of UV origin.  The explicit factorization of the NLL
contributions of UV origin is obtained by means of projector
techniques \cite{Denner:2006jr,nextpaper} since the soft-collinear
approximation is not applicable in this case.  After the
abovementioned minimal UV subtraction, we perform a finite 
renormalization that restores the on-shell normalization 
of the wave functions and translates the couplings to the 
minimal-subtraction 
scheme at the scale $\muR$.

For processes involving heavy quarks, 
in addition to the contributions listed above 
also diagrams with Yukawa interactions contribute.
Many of them are suppressed 
due to the behaviour of the interactions 
of scalar fields in the soft/collinear and $M\to 0$ limits.
Only three types of diagrams are not suppressed,
\unitlength 0.6pt\SetScale{0.6}
\beqar\label{twoloopdiag3}%\refeq{twoloopdiag3}
\vcenter{\hbox{\diagVIIy{$\leg{i}$}{$\leg{j}$}{$\scriptscriptstyle{V}$}{$\scriptscriptstyle{\Phi}$}{\blob}}}
,\quad
\vcenter{\hbox{\diagVy{$\leg{i}$}{$\leg{j}$}{$\scriptscriptstyle{V}$}{$\scriptscriptstyle{\Phi}$}
{\blob}}}
,\quad
\vcenter{\hbox{
\diagIIIy{$\leg{i}$}{$\leg{j}$}{$\scriptscriptstyle{\Phi'}$}{$\scriptscriptstyle{V}$}{$\scriptscriptstyle{\Phi}$}{\blob}}}
.
\eeqar
Moreover it turns out that the sum of the NLL terms resulting from
these diagrams cancels owing to global gauge invariance of
the Yukawa sector \cite{nextpaper}. As a consequence,
Yukawa interactions contribute only through fermionic self-energy diagrams 
that appear in the renormalization of the heavy-quark wave functions.

\section{High-energy expansion of the loop integrals}
The loop integrals have been calculated in the 
$Q^2\gg M^2$ limit to NLL accuracy using 
the sector-decomposition technique \cite{Binoth:2003ak,Denner:2004iz} and, alternatively,
the method of expansion by regions 
\cite{Beneke:1997zp,Smirnov:1998vk,Smirnov:1999bza,Jantzen:2006jv}.
\subsection{Sector decomposition}
Sector decomposition is based on the Feynman-parameter
representation. Exploiting general properties of
soft, collinear and UV singularities, 
one can factorize these singularities in Feynman-parameter
space in such a way that 
the divergent integrations assume a standard form.
If
all particles are massless these singular integrals are trivial and
produce simple $\varepsilon$-poles \cite{Binoth:2003ak}.  In
presence of massive particles, the general solution is available to NLL accuracy
\cite{Denner:2004iz}. This permits to construct,
 in a completely
automatized way,
finite integral
representations for the coefficients of all NLL terms of type
$\veps^{-k}\ln^{j-k}(Q^2/M^2)$.  
The integrands are smooth 
and can easily be integrated numerically.%
\footnote{
This holds only in the unphysical region
where all invariants $r_{ij}$ are negative,
but allows for very
useful checks.}
Moreover, although these integrals do not have a
standard form, it turns out that they are relatively simple. In
practice they can be solved in
analytic form by means of standard tricks (partial fractioning,
integration by parts, etc.) combined with a table of elementary
integrals.  Also this part of the calculation is 
highly automatized, however, 
the set of integration rules might need to be
extended when new diagrams are computed.

\subsection{Expansion by regions}
Within the expansion-by-regions
method~\cite{Beneke:1997zp,Smirnov:1998vk,Smirnov:1999bza,Jantzen:2006jv},
the integration domain of the loop momenta is divided into regions
corresponding to the asymptotic limit considered.  The integrand is
appropriately expanded in every relevant region, and each of the
expanded terms is integrated over the whole integration domain.
Each
expanded term has a unique order in powers of $Q$
and $M$, but the 
on-shell momenta~$p_i$ of massive fermions
involve two scales,
$p_i^2\sim M^2$ and $2p_ip_j\sim Q^2$.
In order to separate these scales, the massive momenta are
reparametrized in terms of light-like momenta~$\tilde p_{i,j}$ as
$p_i = \tilde p_i + p_i^2/(2 \tilde p_i \tilde p_j) \tilde p_j$.
For the loop integrals needed in this calculation, the following
regions for each loop momentum~$k$ are relevant: hard, soft and
ultrasoft regions, where $k^2$ is of the order $Q^2$, $M^2$ and
$M^4/Q^2$, respectively; 
collinear, softcollinear and
ultracollinear regions, where $k$ is collinear to one of the external
momenta and $k^2$ is of the order $M^2$, $M^4/Q^2$ and $M^6/Q^4$,
respectively.  The latter two regions contribute only if the 
external fermions are massive.
The expanded loop integrals have been evaluated using Mellin--Barnes
representations, from which the extraction of NLLs
has been automatized.
In addition, 
Mellin--Barnes representations of the unexpanded integrals
have been used to check the completeness of the set of regions used
for the expansion.

\section{Results and conclusions}
The renormalized 2-loop NLL amplitude
is expressed in terms of 
combinations of SU(2)
matrices acting on the tree-level amplitude.  This expression is
simplified by means of SU(2) algebraic identities and
additional relations resulting from global gauge invariance.
The  final result 
is consistent
with the double-exponentiation formula\footnote{In the following we set $\muD^2=Q^2e^{\gamma_{\mathrm{E}}}/(4\pi)$.}
\beqar\label{eq:exp}%\refeq{eq:exp}
\M
&=& 
\exp
\left(\frac{\alpha}{8\pi}
\sum_{i,j=1\atop j\neq i}^n
\de^\elm_{ij}
\right)
\exp\left(\frac{\alpha}{8\pi}
\sum_{i,j=1\atop j\neq i}^n
\de^\sew_{ij}
\right)
\left(
1+\frac{\alpha}{8\pi}
\sum_{i,j=1\atop j\neq i}^n
\de^Z_{ij}
\right)
\M_0.
\eeqar
The term 
$\de^\sew_{ij}$
behaves as if all electroweak gauge bosons would have the same mass
$M$  and is
symmetric with respect to
SU(2)$\times$U(1) transformations,
\beqar\label{sewfact}%\refeq{sewfact}
\de^\sew_{ij}
&=&
-\sum_{V=B,W^a}I^{\bar V}_i I^V_j
I(\veps,M,-r_{ij})
-
\frac{1}{(4\pi)^2}
\left(
g_1^2 \frac{Y_i Y_j}{4}b^{(1)}_{1}+
g_2^2 T^a_i T^a_j b^{(1)}_{2}
\right)
J(\veps,M,\muR^2)
.\eeqar 
Here  $T^a_i$ and $Y_i$ are the generators of the 
SU(2)$\times$U(1) group,
$g_{1,2}$ are the corresponding coupling constants, and 
$b^{(1)}_{1,2}$ the one-loop $\beta$-function
coefficients  \cite{Denner:2006jr}.
The NLL functions read
\beqar\label{oneloopsew}%\refeq{oneloopsew}
I(\veps,M,r) =
-L^2-\frac23L^3\veps-\frac14L^4\veps^2 +
\left[\frac32-\ln\left(\frac{r}{Q^2}\right)
  -\frac{y_i^{\kappa_i}}{C^\ew_i}\frac{g_2^2\Mt^2}{8e^2\MW^2} \right]
\left(2L+L^2\veps+\frac13L^3\veps^2\right)
\eeqar 
and
$J(\veps,M,\muR^2)
=\left[
I(2\veps,M,Q^2)-
(Q^2/\mu^2_\rR)^\veps I(\veps,M,Q^2)
\right]/\veps$
with $L=\ln(Q^2/M^2)$.
In \refeq{oneloopsew} $C^\ew_i=\sum_{V}I^{\bar V}_i I^V_i$,
and the 
Yukawa terms proportional to $\Mt^2$
contribute with factors $y_{\Pt,\Pb}^{\rL}=1$, $y_\Pt^{\rR}=2$,
$y_\Pb^{\rR}=0$,
for heavy quarks $i=\Pb,\Pt$ with chirality $\kappa_i=\rR,\rL$,
and $y_i^{\kappa}=0$ otherwise.
The term $\de^\elm_{ij}$ in \refeq{eq:exp} depends on the $\gamma$--W mass
gap and contains all soft/collinear singularities due to massless photons.  
These singularities involve single and double $\veps$-poles. Thus the 
terms  $\de^\sew_{ij}$
and $\de^Z_{ij}$ in \refeq{eq:exp} need to be expanded up to $\order(\veps^2)$.
We find that  $\de^\elm_{ij}$ behaves as in QED
\cite{nextpaper}.
Finally we find that all contributions depending on the
Z--W mass difference can be factorized into the one-loop
term $\de^Z_{ij}
=
-I^Z_i I^Z_j
\ln\left({\MZ^2}/{\MW^2}\right)
\left[2 L+2L^2\veps+L^3\veps^2\right]$.
Apart from these latter $\ln(\MZ^2/\MW^2)$ terms, 
all effects  of symmetry breaking (terms proportional to
$\MW/\MZ$, mixing parameters and vacuum expectation value) disappear due to 
non-trivial cancellations between different Feynman diagrams.  These
cancellations originate from relations between the
gauge-boson masses, the mixing parameters, and the 
 vacuum expectation value.
The result is in agreement with the 
resummation prescriptions 
proposed for massless \cite{Kuhn:2000nn,Kuhn:2001hz}
and massive fermions
\cite{Melles:2001ye}
and exhibits strict analogies
with the general form of two-loop singularities in massless QCD
\cite{Catani:1998bh}.

\end{document}